# Impact of the honeycomb spin-lattice on topological magnons and edge states in ferromagnetic 2D skyrmion crystals

Doried Ghader*[1] and Bilal Jabakhanji[1]

[1] College of Engineering and Technology, American University of the Middle East, Egaila 54200, Kuwait

*doried.ghader@aum.edu.kw

## Abstract

Magnons have been intensively studied in two-dimensional (2D) ferromagnetic (FM) skyrmion crystals (SkXs) stabilized on Bravais lattices, particularly triangular and square lattices. In these systems, topological edge states (TESs) have been reported in higher-energy magnon gaps, while the first magnon gap is found to be topologically trivial. In this context, antiferromagnetic (AFM) SkXs on the triangular spin lattice have been considered potentially more interesting for applications, since TESs emerge already in the first magnon gap. Meanwhile, the magnon topology of SkXs stabilized on non-Bravais spin lattices remains largely unexplored. In this work, we theoretically investigate the magnon band structure and TESs in 2D FM SkXs stabilized on the honeycomb spin lattice, including experimentally motivated parameter sets relevant to van der Waals magnets. We show that chiral TESs emerge in the first magnon gap over significant ranges of the Dzyaloshinskii–Moriya interaction and single-ion magnetic anisotropy. Magnetic-field-driven topological phase transitions modify the number of these TESs before eventually trivializing them. In addition, we find that TESs can coexist in the first and higher magnon gaps, which could enable frequency-multiplexed magnonic edge transport. These findings highlight the role of lattice geometry in shaping the magnon topology and edge transport in noncollinear spin textures.

## 1. Introduction

Recent theoretical studies [1–11] have established skyrmion crystals (SkXs) [12–16] as exceptional platforms for exploring topological magnonics and magnon spintronics. In these noncollinear spin textures, magnons (the quanta of spin waves) propagate within an effectively nontrivial magnetic superlattice, experiencing emergent fields generated by the spatially varying magnetization [5,8,17–22]. This scenario profoundly impacts the magnon band structure, resulting in nonzero Berry curvatures, quantized Chern numbers, and topological band gaps. A significant consequence of this nontrivial band topology is the emergence of chiral edge states localized at the boundaries of the SkX [1,4,5,10]. These unidirectional magnon modes facilitate spin and heat transport without backscattering, potentially enabling low-dissipation information transfer. Experimentally, Akazawa et al. provided compelling evidence of a topological thermal Hall signal in the SkX phase of the insulating polar magnet $GaV_4Se_8$, consistent with magnonic edge-state transport [23]. Furthermore, the theoretical prediction of emergent magnon Landau levels was dramatically confirmed through inelastic neutron scattering experiments on $MnSi$ in its skyrmion phase [20]. These experimental findings strongly support theoretical models, directly confirming that noncollinear SkX backgrounds give rise to robust topological magnon band structures.

Considerable effort has been devoted to investigating the magnon band topology and edge states in SkXs on the two-dimensional (2D) triangular Bravais spin lattice [4–7,9,24]. In particular, Néel-type ferromagnetic (FM) SkXs on the triangular lattice have been shown to exhibit gapped magnon bands whose modal characteristics can be identified by examining spin precession patterns at the Brillouin zone (BZ) center [4,5]. Characteristic excitations include the clockwise (CW) and counterclockwise (CCW) modes, corresponding to collective spin precessions around each skyrmion core, as well as the breathing mode, which involves a periodic expansion and contraction of the skyrmion.

Importantly, in these FM SkXs the two lowest-energy magnon bands are topologically trivial [1,4,5,20,21]. Consequently, the first two magnon gaps do not host topological edge states (TESs). In contrast, the third magnon band between the CCW and breathing modes can acquire nontrivial topology within a specific range of low magnetic fields, giving rise to chiral TESs within this gap. Díaz et al. [4] demonstrated that increasing the magnetic field closes and subsequently reopens the third gap, thereby inducing a topological phase transition (TPT). Following this transition, the gap becomes topologically trivial and the associated TESs disappear. These theoretical insights highlight the role of the external magnetic field as an effective tuning parameter, enabling on-demand switching of magnonic edge transport through controlled manipulation of band topology, thus providing a practical route to externally modulate magnonic currents in SkXs.

In addition, Díaz et al. investigated magnon excitations in antiferromagnetic (AFM) SkXs realized on a triangular spin lattice [1]. They showed that AFM SkXs host topological magnon bands already at low energies, with the lowest band carrying a finite Chern number. As a result, chiral magnon edge states emerge within the first bulk magnon gap, which is the lowest energy window where edge modes can appear. Díaz et al. further argued that edge states in this low-energy regime are particularly advantageous, since only small excitation energies are required and both magnon–magnon interactions and linewidth broadening are minimal. As a result, their analysis indicates longer magnon lifetimes and an enhanced robustness of the noninteracting magnon description [25].

The magnon excitation spectrum in 2D SkXs is expected to reflect the underlying spin lattice geometry, analogous to simpler collinear FM cases. A 2D collinear ferromagnet on a triangular Bravais lattice supports a single magnon band, with a bandwidth determined primarily by its coordination number ($z = 6$). By contrast, the honeycomb lattice is a non-Bravais, bipartite structure with two sites per unit cell and a lower coordination number ($z = 3$). Consequently, a collinear ferromagnet on the honeycomb lattice hosts two magnon bands (acoustic and optical), which intersect at Dirac nodal points located at the BZ corners in the absence of symmetry-breaking interactions. The reduced coordination number of the honeycomb lattice decreases the magnon bandwidth relative to that of the triangular lattice. Furthermore, the honeycomb geometry enables richer topological magnon phenomena. For instance, the inclusion of next-nearest-neighbor (NNN) Dzyaloshinskii–Moriya interaction (DMI), permitted by honeycomb-lattice symmetry, opens gaps at the Dirac magnon nodes, resulting in nonzero Chern numbers and chiral magnonic edge states [26–30]. In contrast, a triangular spin lattice collinear ferromagnet, due to having only a single magnon branch, inherently lacks such band-crossing topology and thus cannot support a nonzero Chern invariant or analogous topological edge modes.

Given the profound impact of lattice geometry on the magnon spectrum in 2D collinear FM phases, it is anticipated that the honeycomb lattice geometry in the SkX phases could enable magnonic band topologies and field-driven TPTs beyond those realizable in simpler triangular SkXs. Indeed, several 2D van der Waals magnets (e.g., $CrI_3$, $CrBr_3$, $Cr_2Ge_2Te_6$) naturally exhibit honeycomb spin lattice structures [29,31–40]. Nonetheless, material-specific studies of magnons in 2D SkXs remain scarce, despite their pivotal role at the intersection of two forefront research fields: van der Waals 2D magnetism and skyrmion-based topological magnonics.

In this work, we theoretically investigate magnons in Néel-type FM SkXs on 2D honeycomb spin lattices. We consider several models characterized by different sets of magnetic parameters, including NNN DMI and single-ion magnetic anisotropy (SIMA). Among these, we include experimentally motivated parameter sets corresponding to $CrI_3$ and $CrBr_3$.

Following earlier studies, skyrmions are stabilized through interfacial nearest-neighbor (NN) DMI [1,2,4,5,41]. The SkX is obtained using stochastic Landau–Lifshitz–Gilbert (sLLG) simulations. In addition, we introduce a numerical scheme that systematically tracks the evolution and deformation of the SkX as the external magnetic field is varied.

We find that TESs emerge within the first bulk magnon gap over significant ranges of SIMA and NN DMI. The number of TESs can be modified through magnetic-field-driven topological phase transitions, and the edge states are eventually suppressed beyond a critical magnetic field. We further observe that the range of NN DMI supporting these TESs decreases as the SIMA is reduced, allowing us to identify a lower bound for the SIMA below which TESs no longer emerge. Within this framework, $CrI_3$, which is characterized by a relatively large SIMA, can support such TESs, whereas $CrBr_3$, with its much smaller SIMA, does not exhibit this behavior. We also find that, within the experimentally relevant values explored in this work, the NNN DMI does not have a significant effect on the magnon topology of honeycomb-based SkXs. This behavior stands in sharp contrast to the collinear FM phase, where the NNN DMI plays a crucial role in inducing topological magnon bands. Finally, we demonstrate that TESs can coexist within the first and higher magnon gaps, which could enable frequency-resolved chiral magnon transport.

## 2. Modeling the SkX

We consider a spin Hamiltonian on a 2D honeycomb lattice incorporating FM Heisenberg exchange, SIMA, and both NN and NNN DMIs. The NN DMI is of interfacial origin [1,2,4,5,41], while the NNN DMI arises intrinsically and has been observed experimentally in several honeycomb 2D magnets [27,28,30,42]. The Hamiltonian is given by

$$\mathcal{H} = -J\sum_{\langle i,j\rangle} \boldsymbol{\mathcal{S}}_i \cdot \boldsymbol{\mathcal{S}}_j - \sum_{\langle i,j\rangle} \boldsymbol{d}_{ij} \cdot \boldsymbol{\mathcal{S}}_i \times \boldsymbol{\mathcal{S}}_j - \mathcal{A}\sum_i (\mathcal{S}_i^z)^2 - \sum_{\langle\langle i,j\rangle\rangle} \boldsymbol{D}_{ij} \cdot \boldsymbol{\mathcal{S}}_i \times \boldsymbol{\mathcal{S}}_j - B\sum_i \mathcal{S}_{\boldsymbol{i}}^{\boldsymbol{z}} \tag{1}$$

Here, $\boldsymbol{\mathcal{S}}_i$ denotes the spin operator at site $i$ of the honeycomb lattice. The first three terms describe the FM NN Heisenberg exchange with strength $J$, the interfacial-type NN DMI with vectors $\boldsymbol{d}_{ij}$ ($d = |\boldsymbol{d}_{ij}|$), and the SIMA with strength $\mathcal{A}$. The fourth term represents the intrinsic NNN DMI with strength $D = |\boldsymbol{D}_{ij}|$, while the last term accounts for the Zeeman coupling due to an external magnetic field $B$ applied along the $z$-axis, normal to the honeycomb lattice. The vectors $\boldsymbol{d}_{ij}$ and $\boldsymbol{D}_{ij}$ define the chiral directions of the NN and NNN DMIs, respectively. These vectors are illustrated in Supplementary Fig. S1.

In numerical studies of magnons, it is common to set $J = 1$ and $S = 1$ [1,2,4,5,41]. This choice is equivalent to rescaling the Hamiltonian as $\mathcal{H} \to \mathcal{H}/JS$, such that magnon energies are expressed in units of $JS$, without altering the underlying physics. Under this convention, the intrinsic parameters take the normalized forms $\tilde{J} = 1, \tilde{D} = D/J$, and $\tilde{\mathcal{A}} = \mathcal{A}/J$, and the extrinsic parameters are normalized as $\tilde{d} = d/J$ and $\tilde{B} = B/J$. Note that $\tilde{J}$, $\tilde{D}$, $\tilde{\mathcal{A}}$, and $\tilde{d}$ are unitless, whereas $\tilde{B}$ is in units of $g\mu_B$.

| | **Material** | **Experimental Parameters** | **Normalized Parameters** |
|---|---|---|---|
| **Model 1** | − | − | $\tilde{J} = 1, \tilde{D} = 0, \tilde{\mathcal{A}} = 0.15$ |
| **Model 2** [30] | $CrI_3$ | $J = 2.13\, meV, D = 0.19\, meV, \mathcal{A} = 0.22\, meV$ | $\tilde{J} = 1, \tilde{D} = 0.089, \tilde{\mathcal{A}} = 0.103$ |
| **Model 3** | − | − | $\tilde{J} = 1, \tilde{D} = 0, \tilde{\mathcal{A}} = 0.08$ |
| **Model 4** [28] | $CrI_3$ | $J = 2.11\, meV, D = 0.09\, meV, \mathcal{A} = 0.123\, meV$ | $\tilde{J} = 1, \tilde{D} = 0.043, \tilde{\mathcal{A}} = 0.058$ |
| **Model 5** | − | − | $\tilde{J} = 1, \tilde{D} = 0, \tilde{\mathcal{A}} = 0.045$ |
| **Model 6** [43] | $CrBr_3$ | $J = 1.48\, meV, D = 0, \mathcal{A} = 0.029\, meV$ | $\tilde{J} = 1, \tilde{D} = 0, \tilde{\mathcal{A}} = 0.0196$ |

**Table 1.** Magnetic parameters for Models 1–6. Column 3 lists the intrinsic magnetic parameters, where $J$ denotes the FM Heisenberg exchange, $D$ the NNN DMI, and $\mathcal{A}$ the SIMA. Their normalized counterparts, $\tilde{J} = 1$, $\tilde{D} = D/J$, and $\tilde{\mathcal{A}} = \mathcal{A}/J$, are given in column 4. Models 2, 4, and 6 correspond to experimentally motivated parameter sets for monolayer $CrI_3$ and $CrBr_3$ in their collinear FM phases. Models 1, 3, and 5 are hypothetical and provide interpolating values of $\tilde{\mathcal{A}}$ within the range $0.0196 \leq \tilde{\mathcal{A}} \leq 0.103$ spanned by the experimental models.

We will show that typical values of the NNN DMI reported experimentally for 2D magnets do not affect the magnon topology. Consequently, $\tilde{\mathcal{A}}$ emerges as the only intrinsic parameter that meaningfully distinguishes the models in Eq. 1. We investigate systems with $\tilde{\mathcal{A}} \leq 0.15$ and illustrate our results using six representative models, labeled Models 1–6, with decreasing values of $\tilde{\mathcal{A}}$. The corresponding magnetic parameters are summarized in Table 1. Among these, Models 2, 4, and 6 are adopted from experimental studies of magnons in the collinear FM phase of monolayer $CrI_3$ [28,30] and $CrBr_3$ [43].

We simulated the ground state using the sLLG equations within the Vampire software package [44]. FM (Néel-type) SkXs were observed above a threshold NN DMI, which increases with $\tilde{\mathcal{A}}$. The simulations were initialized from random spin configurations at high temperatures and then gradually cooled to near zero temperature. At minimal field ($\tilde{B}_{min}$), the skyrmions are densely packed, forming a triangular lattice. It should be noted, however, that the SkXs generated by Vampire are not perfectly ordered due to the random nucleation of NN DMI-induced skyrmions [45–47]. Nevertheless, idealized SkX configurations can be generated based on the Vampire results using suitable analytical functions [6]. Once the SkX is stabilized, the magnetic field $\tilde{B}$ is incrementally increased to study the evolution of the texture, with the temperature maintained near zero. As $\tilde{B}$ increases, individual skyrmions gradually shrink in size while remaining pinned to their original lattice sites, thus preserving the triangular SkX structure and its BZ. Eventually, beyond a critical magnetic field $\tilde{B}_{max}$, the skyrmions are gradually annihilated, and the system gradually transitions into a uniform ferromagnetic state.

Determining the skyrmion size as a function of the magnetic field is essential for understanding its impact on magnon excitations. We develop a practical method to extract the field-dependent effective skyrmion width from the integrated out-of-plane spin density of a SkX. At the minimal magnetic field, the densely packed SkX features hexagonal skyrmions with initial width $w_0$ and area $A_0 = \frac{\sqrt{3}}{2} w_0^2$ (see Supplementary Fig. S2(a)). Notably, $w_0$ defines the inter-skyrmion distance and hence matches the SkX periodicity (Supplementary Fig. S2(c)). Let $S_0^z(\boldsymbol{r})$ denote the interpolated z-component of the spin field in this configuration. The corresponding out-of-plane spin density is

$$\eta_0 = \frac{1}{A_0} \iint_{A_0} S_0^z(\boldsymbol{r}) ds \tag{2a}$$

At an increased magnetic field $\tilde{B} > \tilde{B}_{min}$, each skyrmion shrinks and no longer occupies the entire unit cell (Supplementary Fig. S2(b)). Assuming a self-similar deformation, the skyrmion at this field has width $w < w_0$ and area $A_s = \frac{\sqrt{3}}{2} w^2 < A_0$, while the remaining region corresponds to spins aligned along the z-direction. The new out-of-plane spin density is given by

$$\eta = \frac{1}{A_0} \iint_{A_0} S^z(\boldsymbol{r}) ds \tag{2b}$$

where $S^z(\boldsymbol{r})$ is the updated interpolation function at field $\tilde{B}$. Recognizing that $\iint_{A_s} S^z(\boldsymbol{r}) ds = \eta_0 A_s$, the updated density becomes

$$\eta = \frac{1}{A_0} \iint_{A_0 - A_s} 1\, ds + \frac{1}{A_0} \iint_{A_s} S^z(\boldsymbol{r}) ds = 1 + \frac{A_s}{A_0} (\eta_0 - 1) \tag{2c}$$

Substituting $\frac{A_s}{A_0} = \frac{w^2}{w_0^2}$, we obtain the skyrmion width as a function of $\eta$

$$w = w_0 \sqrt{\frac{\eta - 1}{\eta_0 - 1}} \tag{2d}$$

In practice, we used *Mathematica* to analyze the spin textures obtained from Vampire sLLG simulations and numerically extracted $w(\tilde{B})$ using Eqs. (2a), (2b), and (2d). A representative example is shown in Fig. 1 for $CrI_3$ (Model 2) with $\tilde{d} = 0.45$.

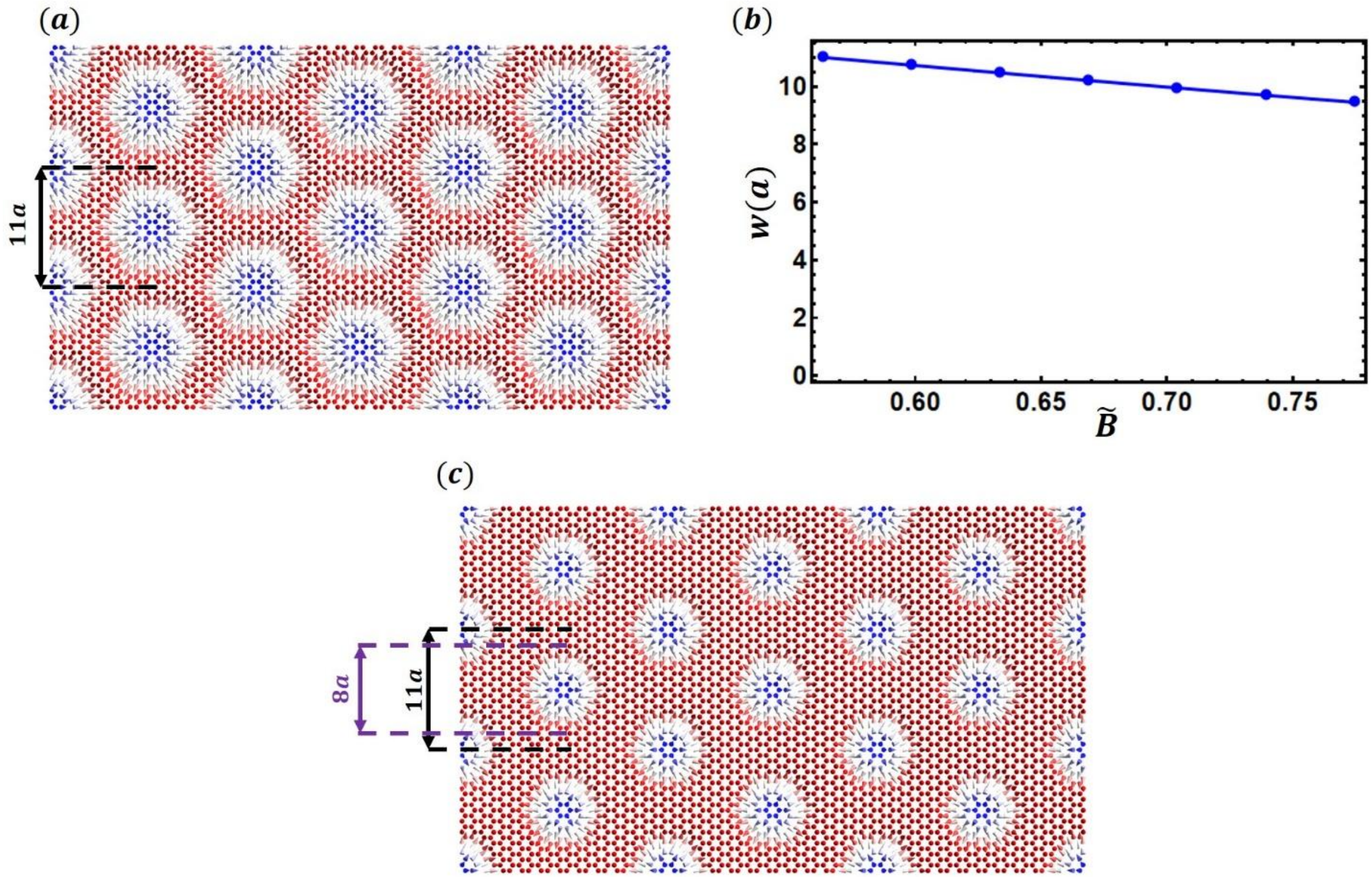

**Figure 1.** Representative SkX configurations on the honeycomb spin lattice for $CrI_3$ (Model 2) at $\tilde{d} = 0.45$. (a) Densely packed SkX at the minimal magnetic field $\tilde{B}_{min} = 0.55$, corresponding to a skyrmion width $w_0 = 11a$. The color map represents the z-component of the normalized spin, ranging from −1 (blue) to 1 (red). (b) Skyrmion width $w$ (in units of $a$) as a function of the magnetic field. The dots are extracted from sLLG simulations using Eq. (2d), while the solid line is a plot of an interpolation function. (c) SkX configuration at $\tilde{B} \approx 1.02$, where the skyrmions shown in panel (a) have shrunk to a width $w \approx 8a$.

## 3. Magnon Hamiltonian for honeycomb-based FM SkXs

We adopt a discrete Holstein–Primakoff bosonization scheme [1–3,5,48] to quantize the spin excitations, which is more suitable than continuum approaches [6] for the relatively small skyrmions appearing in the models under study. While this method has been extensively developed for triangular spin lattice SkXs, we extend it here to treat SkXs on the two-sublattice honeycomb structure and include contributions arising from the NNN DMI. A brief overview of the theoretical framework is provided below, and full technical details can be found in the Supplementary Material.

In the noncollinear SkX state, each spin at lattice site $i$ has a local orientation $\boldsymbol{n}_i$ that differs from the global $z$-axis. To account for this, we perform a local spin-axis rotation at each site [49] to align the local spin quantization axis with the spin's equilibrium orientation $\boldsymbol{n}_i$. This transformation maps the SkX onto an equivalent ferromagnetic state in the rotating frame. We then apply Holstein–Primakoff bosonization in this rotated frame, expressing the spin operators in terms of bosonic magnon creation ($a_i^+$, $b_j^+$) and annihilation ($a_i$, $b_j$) operators defined about the local ordered spin. Here, $a_i$ and $b_j$ denote magnon operators on sublattices A and B, respectively.

The magnetic unit cell of the SkX is defined as the cluster of spins forming a single skyrmion in the minimal-field configuration. For a skyrmion of width $w_0$, this unit cell contains $N_s = w_0^2$ spins per sublattice (i.e., $2N_s$ spins in total). The skyrmion centers form a triangular Bravais lattice in real space, and thus, in momentum space, one can define a magnon Bloch wavevector (or momentum) $\boldsymbol{k}$ in the SkX BZ, which is a mini-BZ relative to that of the atomic lattice.

After Fourier transforming the bosonic operators to momentum space, the quadratic magnon Hamiltonian takes the form,

$$\mathcal{H} = \frac{1}{2}\sum_{\boldsymbol{k}} \Psi^\dagger \, h(\boldsymbol{k}) \, \Psi \tag{3a}$$

with

$$h(\boldsymbol{k}) = \begin{pmatrix} X(\boldsymbol{k}) & Y(\boldsymbol{k}) \\ Y^\dagger(\boldsymbol{k}) & X^T(-\boldsymbol{k}) \end{pmatrix} \tag{3b}$$

and

$$\Psi^\dagger = \begin{pmatrix} a_{\boldsymbol{k}1}^+ & \cdots & a_{\boldsymbol{k}N_s}^+ & b_{\boldsymbol{k}1}^+ & \cdots & b_{\boldsymbol{k}N_s}^+ & a_{-\boldsymbol{k}1} & \cdots & a_{-\boldsymbol{k}N_s} & b_{-\boldsymbol{k}1} & \cdots & b_{-\boldsymbol{k}N_s} \end{pmatrix} \tag{3c}$$

In Eqs. (3), $\boldsymbol{k}$ is a wavevector in the SkX BZ, and $h(\boldsymbol{k})$ is a $4N_s \times 4N_s$ matrix. The block matrices $X(\boldsymbol{k})$ and $Y(\boldsymbol{k})$ are each $2N_s \times 2N_s$, and they encode the exchange, anisotropy, and DM interactions. The explicit forms of these matrices are given in the Supplementary Material.

The matrix $h(\boldsymbol{k})$ is diagonalized using a standard bosonic Bogoliubov transformation following Colpa's method [50]. This procedure yields the magnon band energies $E_n$ $(n = 1, 2, \ldots, 2N_s)$ and corresponding eigenvectors, from which one can compute the Berry curvature and Chern number associated with each band. In particular, the Berry curvatures and Chern numbers are calculated using the numerical method developed by Fukui *et al.* [51]. Extensive details on the implementation of this method for magnonic band structures can be found in previous works [6,52,53], and are omitted here for brevity.

## 4. Magnon bands, topology, and edge states

### 4.1 TESs in the first energy gap

We first consider Model 1, which is characterized by a relatively large magnetic anisotropy $\tilde{\mathcal{A}} = 0.15$ (Table 1). Consequently, SkXs in this model are stabilized only when the NN DMI exceeds a significant threshold value, $\tilde{d}_{\text{th}} = 0.55$. We find that this model hosts TESs in the lowest energy gap within the range $0.55 \leq \tilde{d} \leq 1.2$.

We begin with $\tilde{d} = 1.2$, for which the minimum magnetic field required to stabilize the SkX is $\tilde{B}_{\text{min}} \approx 5.4$. This field yields a densely packed SkX with skyrmion width $w_0 = 5a$. We label the magnon bands by $E_n$ and their associated Chern numbers by $C_n$, with $n = 1, 2, \dots, 2N_s$ in increasing energy order. The two lowest magnon bands at $\tilde{B}_{\text{min}}$ are shown in Fig. 2(a), plotted along the high-symmetry directions of the SkX BZ. Bands $E_1$ and $E_2$ are topological, carrying Chern numbers $\pm 1$, and define the topological phase $P_1 = \{1, -1\}$.

To characterize these modes, we analyze the time evolution of the real-space out-of-plane magnetization for bands $E_1$ and $E_2$ at the BZ center. We emphasize that, in the noncollinear SkX phase, the relevant magnon excitations correspond to collective skyrmion modes, such as CW, CCW, elliptical distortion (ED), and breathing modes, rather than the optical and acoustic branches characteristic of collinear honeycomb ferromagnets. We find that the lowest-energy band $E_1$ corresponds to an ED mode, while the second band $E_2$ exhibits a CCW rotation mode. Notably, these modal characteristics are inverted relative to the triangular-lattice SkX, where $E_1$ is a CCW mode and $E_2$ an ED mode [4].

According to the bulk–edge correspondence, chiral TESs emerge within a given gap when the total Chern number of all bands below that gap is nonzero. In a system finite in both spatial directions, the number of edge states equals this total Chern number, while it is doubled in a geometry that is finite in one direction and infinite in the other. Since $C_1 = 1$ in phase $P_1$, two TESs appear in the first bulk magnon gap in a semi-infinite sample. These states are shown in Fig. 3(a) at $\tilde{B}_{\min}$, computed using a strip geometry that is infinite along the $y$ direction and finite along the $x$ direction. The edge states, highlighted in red and blue, are chiral and propagate unidirectionally along opposite edges of the strip.

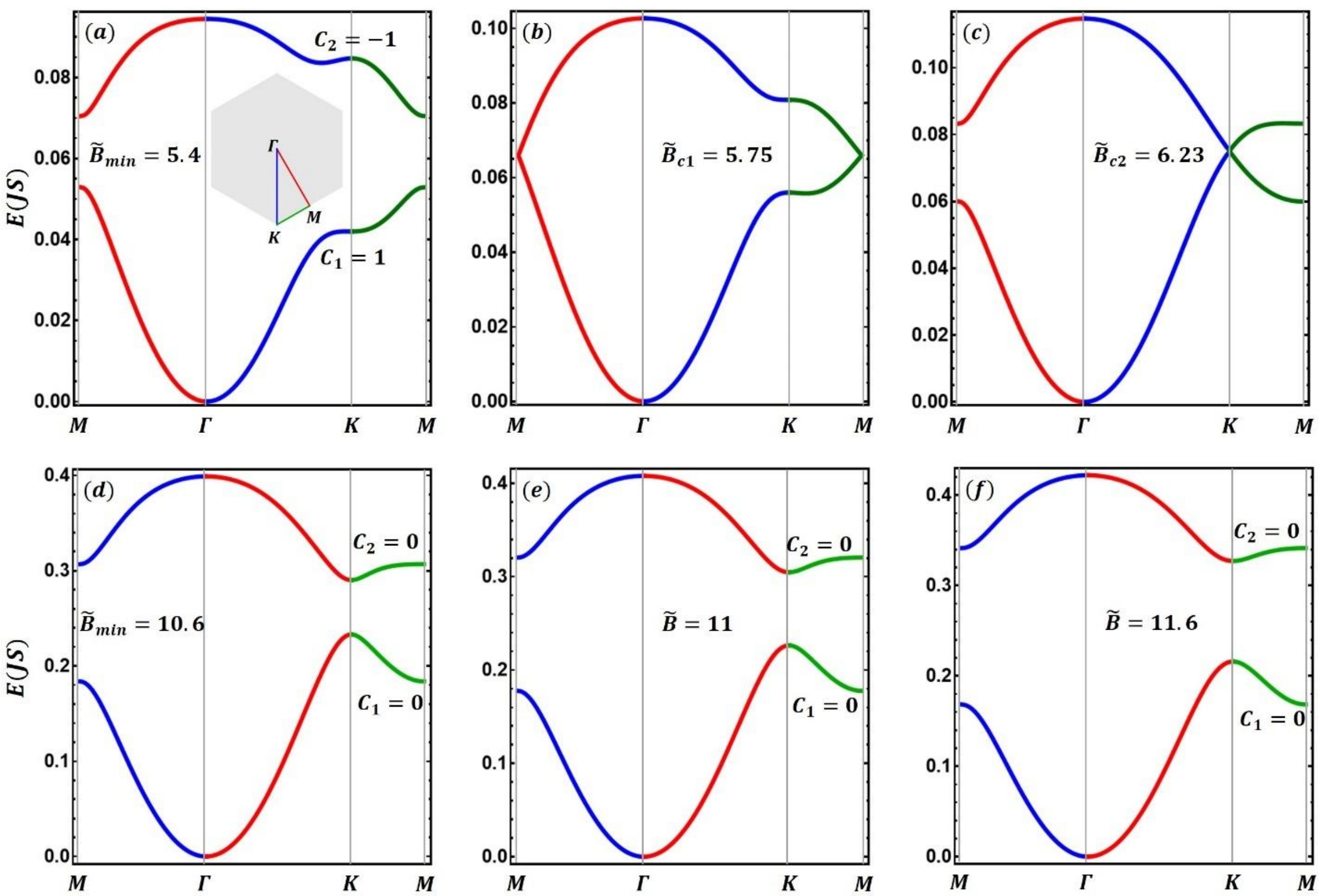

**Figure 2.** Lowest two magnon energy bands $E_1$ and $E_2$ for $\tilde{\mathcal{A}} = 0.15$ and $\tilde{d} = 1.2$ in SkXs on the honeycomb (a–c) and triangular (d–f) spin lattices. The bands are plotted along the high-symmetry directions of the SkX BZ, as indicated in (a). The skyrmion periodicities are $w_0 = 5a$ and $8a$ for the honeycomb and triangular lattices, respectively. In the honeycomb lattice, the bands are topological at $\tilde{B}_{\min}$ (a) and undergo two topological phase transitions at $\tilde{B}_{c1}$ (b) and $\tilde{B}_{c2}$ (c). By contrast, in the triangular lattice the bands are topologically trivial at $\tilde{B}_{\min}$ (d), and the gap between them (the first magnon gap) increases monotonically with the magnetic field (e, f).

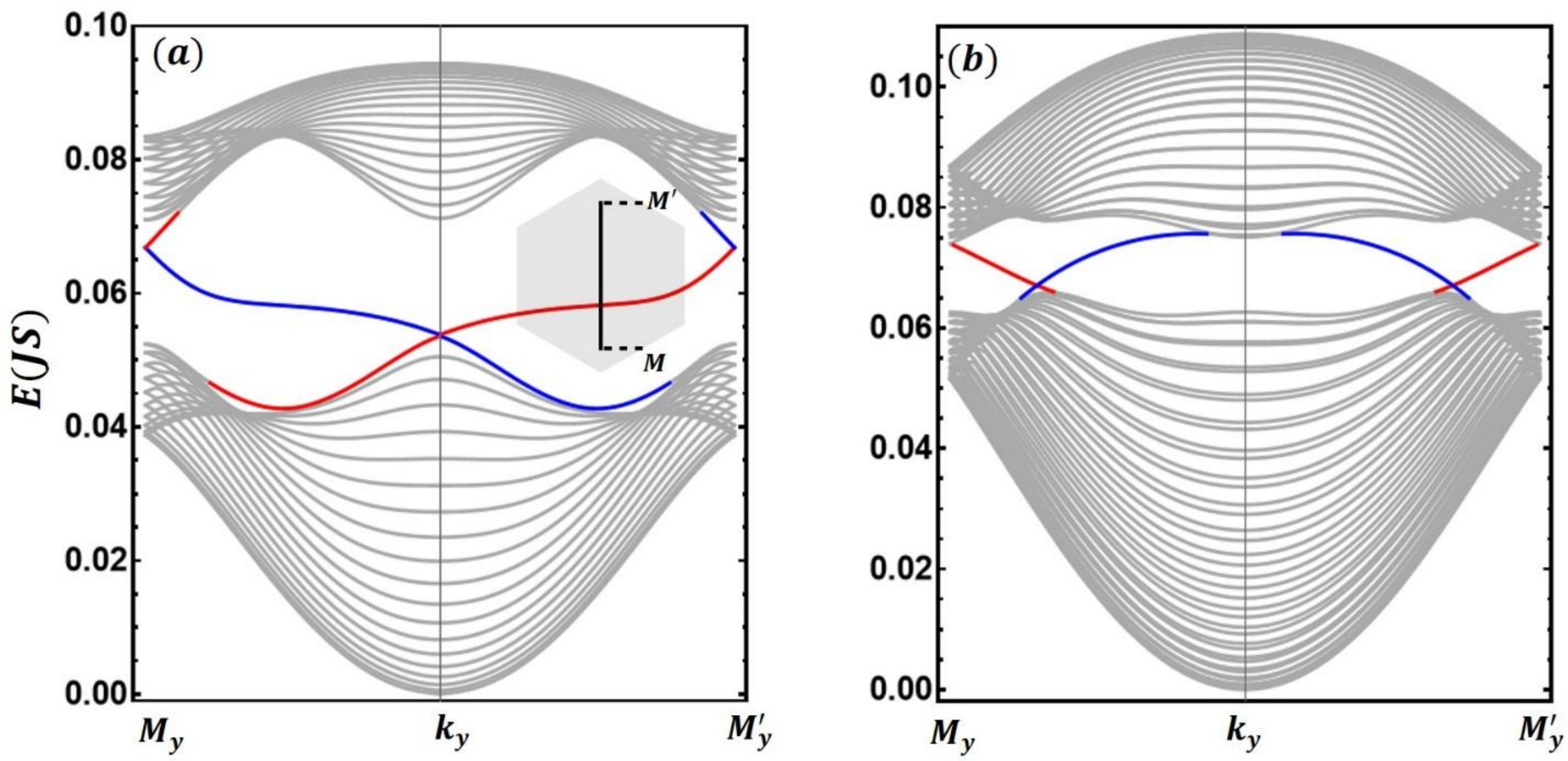

**Figure 3.** Chiral TESs, shown as red and blue curves, in the first magnon gap for the honeycomb-based SkX with $\tilde{\mathcal{A}} = 0.15$ and $\tilde{d} = 1.2$. Panels (a) and (b) correspond to the topological phases $P_1 = \{1, -1\}$ and $P_2 = \{-2,2\}$, respectively. The gray curves denote the bulk magnon bands. The magnetic fields are $\tilde{B}_{\min} = 5.4$ in (a) and $\tilde{B} = 6.03$ in (b). The field in (b) lies between the two critical values $\tilde{B}_{c1} = 5.75$ and $\tilde{B}_{c2} = 6.23$. The spectra are computed using a strip geometry that is infinite along the $y$ direction and finite along the $x$ direction. Accordingly, the dispersions are plotted as functions of $k_y$ along the interval $M_y \leq k_y \leq M_y'$, where $M_y$ and $M_y'$ are the projections of the high-symmetry points $M$ and $M'$ of the BZ, as illustrated in the inset of panel (a).

Increasing the magnetic field closes the first gap at the $M$ point of the BZ at a critical field $\tilde{B}_{c1} \approx 5.75$ (Fig. 2(b)). Upon reopening, this gap closure induces a topological phase transition from $P_1 = \{1, -1\}$ to $P_2 = \{-2,2\}$. Since $C_1 = -2$ in phase $P_2$, the number of chiral TESs in the first gap is doubled. The resulting four edge states are shown in Fig. 3(b). Importantly, this transition does not invert the modal characters, and $E_1$ and $E_2$ remain ED and CCW modes, respectively, after the gap reopens.

A further increase of the magnetic field closes the first gap again at a second critical field $\tilde{B}_{c2} \approx 6.23$, this time at the $K$ point of the BZ (Fig. 2(c)). This second gap closing drives a transition from $P_2 = \{-2,2\}$ to the trivial phase $P_3 = \{0,0\}$. After the gap reopens, the modal characters are inverted, with $E_1$ becoming a CCW mode and $E_2$ an ED mode, and the TESs in the first gap disappear. For $\tilde{B} > \tilde{B}_{c2}$, the first gap increases monotonically, and TESs do not reappear in this gap.

For comparison, Figs. 2(d)–2(f) show the trivial magnon bands $E_1$ and $E_2$ for a triangular-lattice SkX with the same parameters $\tilde{\mathcal{A}} = 0.15$ and $\tilde{d} = 1.2$. In this case, the modal characters are inverted relative to the honeycomb lattice, with $E_1$ and $E_2$ corresponding to CCW and ED modes, respectively, and the bands are topologically trivial at $\tilde{B}_{\min}$ (Fig. 2(d)). Moreover, increasing the magnetic field monotonically enlarges the first gap (Figs. 2(e) and 2(f)), so that no topological phase transitions occur and the gap remains trivial for all $\tilde{B}$. We also note that $\tilde{B}_{\min} \approx 10.6$ and $w_0 = 8a$ in the triangular case, in contrast to $\tilde{B}_{\min} \approx 5.4$ and $w_0 = 5a$ for the honeycomb case. As discussed in Sec. 5, these differences originate from the distinct lattice geometries and interaction networks of the two spin lattices.

We next examine lower values of $\tilde{d}$ within Model 1. TESs persist in the first magnon gap throughout the range $0.55 \leq \tilde{d} \leq 1.2$, except at isolated values of $\tilde{d}$ discussed in Sec. 5. Depending on $\tilde{d}$, bands $E_1$ and $E_2$ may undergo either two successive TPTs, $P_1 \rightarrow P_2 \rightarrow P_3$, or a single transition $P_2 \rightarrow P_3$, as the magnetic field is increased from $\tilde{B}_{\min}$. In the latter case, the system already resides in phase $P_2$ at $\tilde{B}_{\min}$. The evolution of the modal characters within these phases, described above for $\tilde{d} = 1.2$, remains valid throughout this range. Representative examples of TESs in the first gap are shown in Fig. 4 for $\tilde{d} = 0.9$, 0.65, and 0.55, with the corresponding values of $w_0$, $\tilde{B}_{c1}$, and $\tilde{B}_{c2}$ listed in Table 2. The corresponding trivial bands for the triangular lattice are shown in Supplementary Fig. S3.

We now investigate the effect of reducing $\tilde{\mathcal{A}}$ on the edge states in the first magnon gap using Models 2–6. Models 2–5 exhibit the same qualitative behavior as Model 1, including the sequence of topological phases, critical fields, and modal characters. The main difference is that the NN DMI range supporting first-gap edge states, denoted $R_{\mathrm{DMI}}$, shrinks systematically as $\tilde{\mathcal{A}}$ decreases from Model 1 to Model 5. For each model, $R_{\mathrm{DMI}}$ extends from the threshold value $\tilde{d}_{\mathrm{th}}$, required to stabilize the SkX, to an upper bound $\tilde{d}_{\max}$ beyond which first-gap TESs are no longer observed. Both quantities depend on $\tilde{\mathcal{A}}$ and are listed in Table 3. Figure 5 illustrates this trend, showing a clear monotonic decrease of $R_{\mathrm{DMI}}$ with decreasing $\tilde{\mathcal{A}}$. For $\tilde{\mathcal{A}} = 0.045$ (Model 5), we find $\tilde{d}_{\mathrm{th}} \approx 0.31$ and $\tilde{d}_{\max} \approx 0.34$, resulting in a narrow interval $R_{\mathrm{DMI}} \approx 0.03$. This value therefore provides a reasonable estimate for the minimal anisotropy required to generate TESs in the first magnon gap. Consistent with this conclusion, such edge states are absent in Model 6, which corresponds to the $CrBr_3$ parameter set with $\tilde{\mathcal{A}} = 0.0196$.

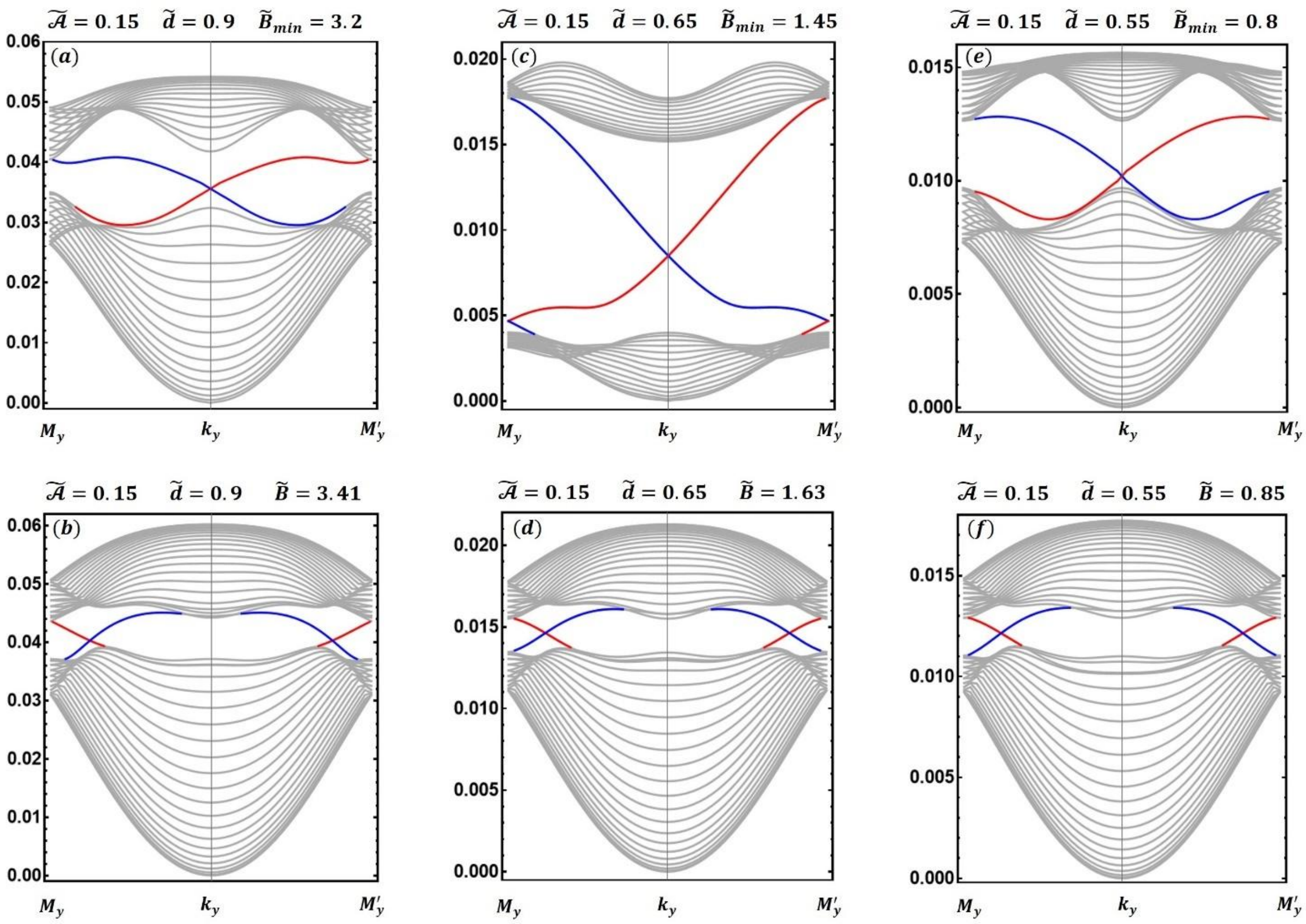

**Figure 4.** Chiral TESs, shown as red and blue curves, in the first magnon gap for honeycomb-based SkXs with $\tilde{\mathcal{A}} = 0.15$. Panels (a), (c), and (e) correspond to phase $P_1 = \{1, -1\}$, while panels (b), (d), and (f) correspond to phase $P_2 = \{-2,2\}$. The results are shown for NN DMI values $\tilde{d} = 0.9$[(a), (b)], 0.65[(c), (d)], and 0.55[(e), (f)]. The gray curves represent the bulk magnon bands. Both bulk and edge states are computed along the $k_y$ segment illustrated in Fig. 3(a). The spectra in phase $P_1$ are evaluated at magnetic fields satisfying $\tilde{B}_{\min} < \tilde{B} < \tilde{B}_{c1}$, while those in phase $P_2$ correspond to fields in the interval $\tilde{B}_{c1} < \tilde{B} < \tilde{B}_{c2}$. The values of $\tilde{B}_{\min}$ are indicated in the panels, while a complete list of the relevant parameters $(\tilde{d}, w_0, \tilde{B}_{\min}, \tilde{B}_{c1}, \tilde{B}_{c2})$ is provided in Table 2.

| $\tilde{d}$ | $w_0$ | $\tilde{B}_{min}$ | $\tilde{B}_{c1}$ | $\tilde{B}_{c2}$ |
|---|---|---|---|---|
| 0.9 | $6a$ | 3.2 | 3.28 | 3.51 |
| 0.65 | $8a$ | 1.45 | 1.6 | 1.66 |
| 0.55 | $9a$ | 0.87 | 0.93 | 0.97 |

**Table 2.** SkX periodicity $w_0$, minimal magnetic field $\tilde{B}_{\min}$, and critical magnetic fields $\tilde{B}_{c1}$ and $\tilde{B}_{c2}$ for honeycomb-based SkXs with $\tilde{\mathcal{A}} = 0.15$ (Model 1) and NN DMI values $\tilde{d} = 0.9$, 0.65, and 0.55. The critical field $\tilde{B}_{c1}$ separates the topological phases $P_1 = \{1, -1\}$ and $P_2 = \{-2,2\}$, while $\tilde{B}_{c2}$ separates phases $P_2 = \{-2,2\}$ and $P_3 = \{0,0\}$ (see main text for details). The corresponding TESs in phases $P_1$ and $P_2$ are shown in Fig. 4.

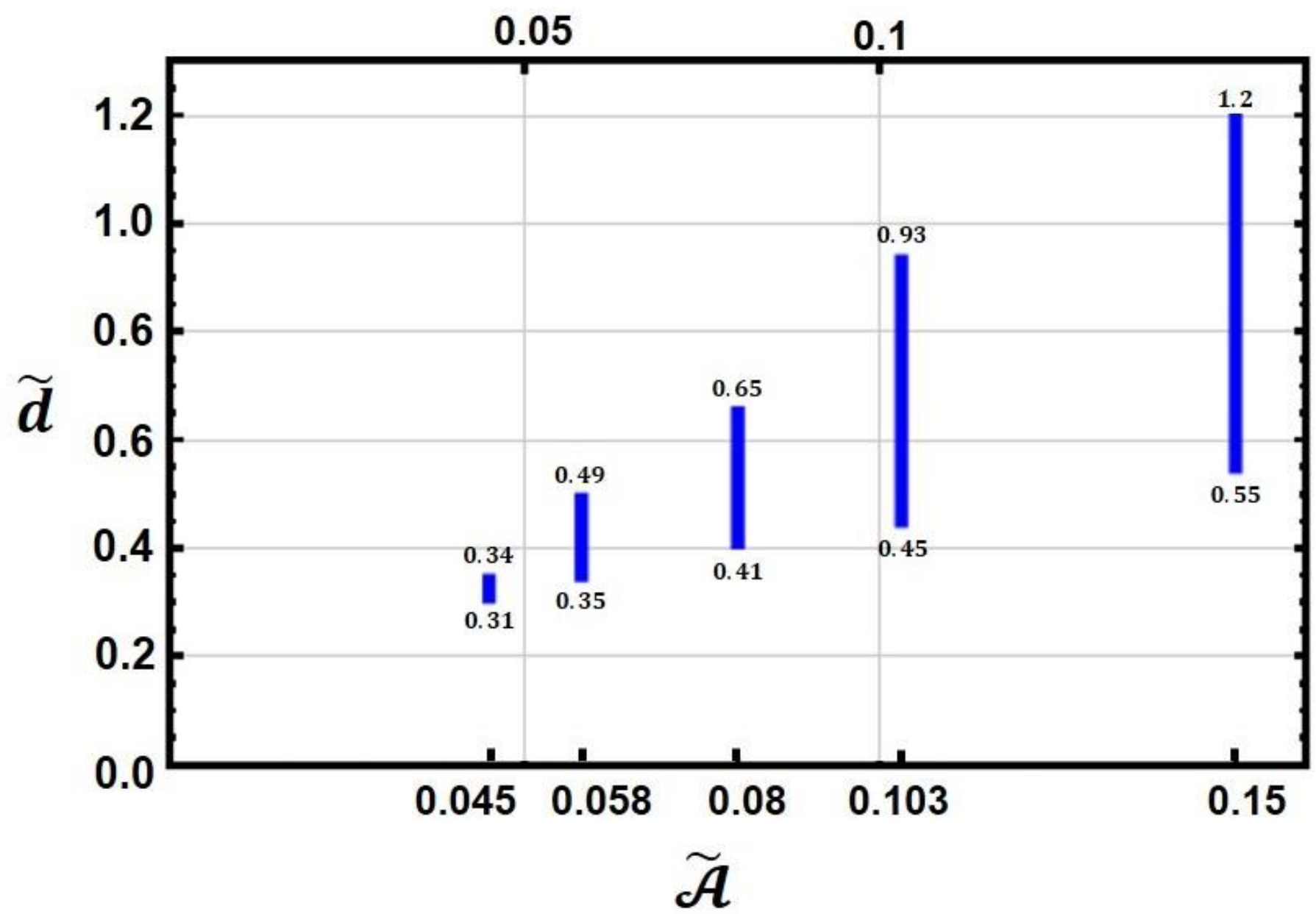

**Figure 5.** NN DMI window $R_{\mathrm{DMI}} = [\tilde{d}_{\mathrm{th}}, \tilde{d}_{\max}]$ over which TESs appear in the first magnon gap, shown as a function of the normalized anisotropy $\tilde{\mathcal{A}}$ for Models 1–5. For each model, the interval $R_{\mathrm{DMI}}$ (blue bar) is bounded by the threshold NN DMI $\tilde{d}_{\mathrm{th}}$ required to stabilize the SkX and the upper limit $\tilde{d}_{\max}$, beyond which first-gap topological edge states are no longer observed. The values of $\tilde{d}_{\mathrm{th}}$ and $\tilde{d}_{\max}$ are indicated below and above each bar, respectively, and are also listed in Table 3.

Examples of TESs in the first magnon gap for Models 2–5 at $\tilde{d}_{\mathrm{th}}$ are shown in Figs. 6 and 7 for phases $P_1$ and $P_2$, respectively. As already demonstrated for Model 1 (Figs. 3 and 4), edge states at larger values of $\tilde{d}$ display the same qualitative features and are therefore omitted for brevity. The corresponding values of $w_0$, $\tilde{B}_{c1}$, and $\tilde{B}_{c2}$ are summarized in Table 3. We note that phase $P_1$ is absent in Model 5 at $\tilde{d}_{\mathrm{th}}$, and the system already resides in phase $P_2$ at $\tilde{B}_{\min}$ (Fig. 7(d)).

Finally, we comment on the role of the NNN DMI $\widetilde{D}$. The results for the $CrI_3$-based models (Models 2 and 4) shown in Figs. 5–7 were obtained using experimentally reported values of $\widetilde{D}$ (Table 1). Setting $\widetilde{D} = 0$, however, leaves the edge states and TPTs unchanged. While the NNN DMI is known to play a crucial role in the collinear FM phase, we find that moderate values of $\widetilde{D}$, such as those reported experimentally for $CrI_3$, do not affect the magnon topology or the TESs in the SkX phase.

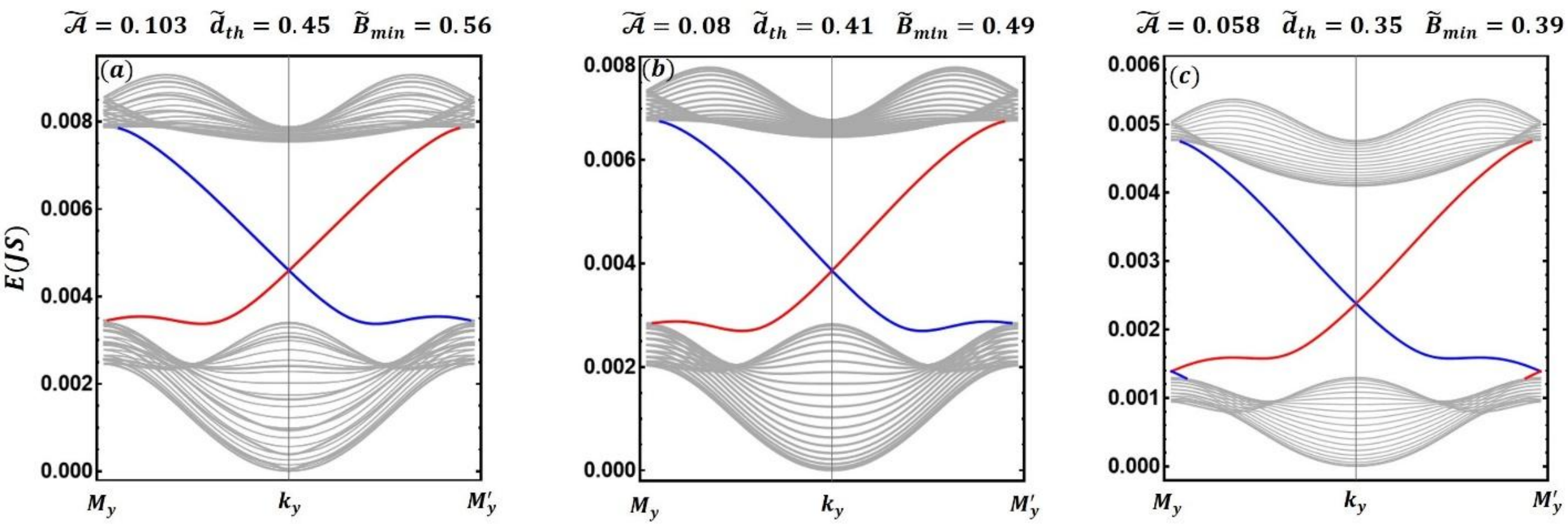

**Figure 6.** Chiral TESs, shown as red and blue curves, in the first magnon gap for phase $P_1 = \{1, -1\}$ at the threshold NN DMI $\tilde{d}_{\text{th}}$. Panels (a)–(c) correspond to Model 2 ($\tilde{\mathcal{A}} = 0.103$), Model 3 ($\tilde{\mathcal{A}} = 0.08$), and Model 4 ($\tilde{\mathcal{A}} = 0.058$), respectively. This phase is absent in Model 5 ($\tilde{\mathcal{A}} = 0.045$). The gray curves represent the bulk magnon bands. Bulk and edge states are computed along the $k_y$ segment illustrated in Fig. 3(a). The spectra are evaluated at the minimal magnetic fields $\tilde{B}_{\text{min}}$, as indicated in the panels. A complete list of the parameters relevant to these results is provided in Table 3.

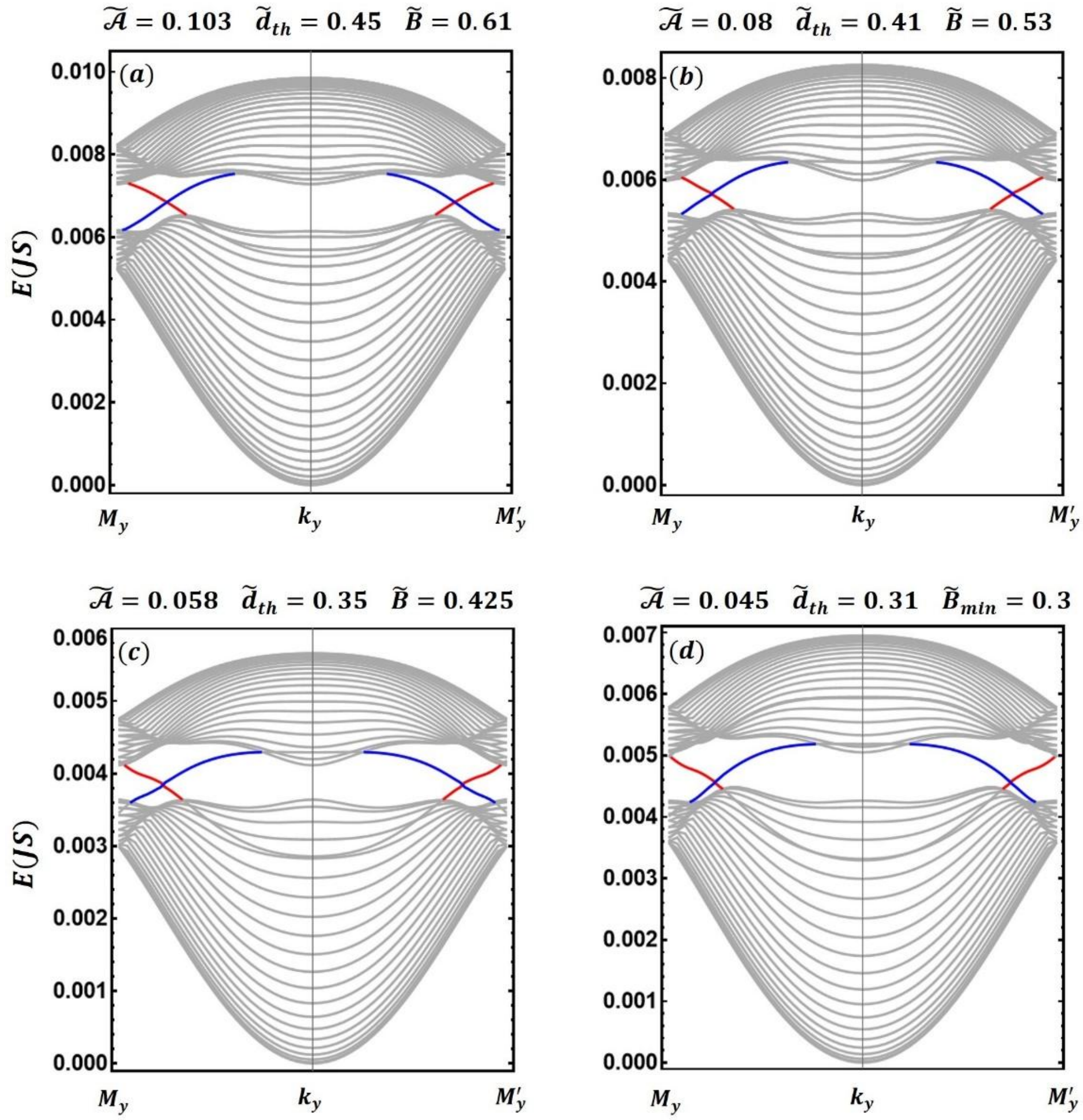

**Figure 7.** Chiral TESs, shown as red and blue curves, in the first magnon gap for phase $P_2 = \{-2,2\}$ at the threshold NN DMI $\tilde{d}_{\text{th}}$. Panels (a)–(d) correspond to Model 2 ($\tilde{\mathcal{A}} = 0.103$), Model 3 ($\tilde{\mathcal{A}} = 0.08$), Model 4 ($\tilde{\mathcal{A}} = 0.058$), and Model 5 ($\tilde{\mathcal{A}} = 0.045$), respectively. The gray curves represent the bulk magnon bands. Bulk and edge states are computed along the $k_y$ segment illustrated in Fig. 3(a). The magnetic fields used for each panel are indicated in the figures. A complete list of the parameters relevant to these results is provided in Table 3.

|  |  |  |  |  |  |  |  |  |
|---|---|---|---|---|---|---|---|---|
| Model 1, $\tilde{\mathcal{A}} = 0.15$ | $\tilde{d}_{th} = 0.55$ | | | | $\tilde{d}_{max} = 1.2$ | | | |
| | $w_0$ | $\tilde{B}_{min}$ | $\tilde{B}_{c1}$ | $\tilde{B}_{c2}$ | $w_0$ | $\tilde{B}_{min}$ | $\tilde{B}_{c1}$ | $\tilde{B}_{c2}$ |
| | $9a$ | 0.87 | 0.93 | 0.97 | $5a$ | 5.4 | 5.75 | 6.23 |
| Model 2, $\tilde{\mathcal{A}} = 0.103$ | $\tilde{d}_{th} = 0.45$ | | | | $\tilde{d}_{max} = 0.93$ | | | |
| | $w_0$ | $\tilde{B}_{min}$ | $\tilde{B}_{c1}$ | $\tilde{B}_{c2}$ | $w_0$ | $\tilde{B}_{min}$ | $\tilde{B}_{c1}$ | $\tilde{B}_{c2}$ |
| | $11a$ | 0.55 | 0.6 | 0.62 | $6a$ | 3.38 | 3.75 | 4 |
| Model 3, $\tilde{\mathcal{A}} = 0.08$ | $\tilde{d}_{th} = 0.41$ | | | | $\tilde{d}_{max} = 0.65$ | | | |
| | $w_0$ | $\tilde{B}_{min}$ | $\tilde{B}_{c1}$ | $\tilde{B}_{c2}$ | $w_0$ | $\tilde{B}_{min}$ | $\tilde{B}_{c1}$ | $\tilde{B}_{c2}$ |
| | $12a$ | 0.49 | 0.52 | 0.54 | $8a$ | 1.65 | 1.83 | 1.92 |
| Model 4, $\tilde{\mathcal{A}} = 0.058$ | $\tilde{d}_{th} = 0.35$ | | | | $\tilde{d}_{max} = 0.49$ | | | |
| | $w_0$ | $\tilde{B}_{min}$ | $\tilde{B}_{c1}$ | $\tilde{B}_{c2}$ | $w_0$ | $\tilde{B}_{min}$ | $\tilde{B}_{c1}$ | $\tilde{B}_{c2}$ |
| | $14a$ | 0.39 | 0.42 | 0.435 | $10a$ | 0.924 | 0.967 | 1.02 |
| Model 5, $\tilde{\mathcal{A}} = 0.045$ | $\tilde{d}_{th} = 0.31$ | | | | $\tilde{d}_{max} = 0.34$ | | | |
| | $w_0$ | $\tilde{B}_{min}$ | $\tilde{B}_{c1}$ | $\tilde{B}_{c2}$ | $w_0$ | $\tilde{B}_{min}$ | $\tilde{B}_{c1}$ | $\tilde{B}_{c2}$ |
| | $15a$ | 0.3 | – | 0.312 | $14a$ | 0.425 | 0.435 | 0.455 |

**Table 3.** Summary of relevant parameters for Models 1–5. The threshold NN DMI $\tilde{d}_{\text{th}}$ denotes the minimum value required to stabilize the SkX. For each model, TESs occur in the first magnon gap within the NN DMI interval $[\tilde{d}_{\text{th}}, \tilde{d}_{\text{max}}]$. The skyrmion periodicity $w_0$, minimal magnetic field $\tilde{B}_{\text{min}}$, and critical magnetic fields $\tilde{B}_{c1}$ and $\tilde{B}_{c2}$ are listed at both $\tilde{d}_{\text{th}}$ and $\tilde{d}_{\text{max}}$. For Model 1, the entries at $\tilde{d}_{\text{th}} = 0.55$ are reproduced from Table 2 to facilitate direct comparison across all models.

## 4.2 Frequency-multiplexed magnonic edge states

As discussed in Secs. 3 and 4.1, a SkX with periodicity $w_0$ supports $2N_s$ magnon bands $E_n$ $(n = 1, 2, \dots, 2N_s)$, where $N_s = w_0^2$, and consequently $2N_s - 1$ magnon gaps. Many of these bands can be topological, giving rise to TESs in several gaps. Moreover, when a large number of bands is considered, magnetic fields are expected to induce multiple TPTs. A comprehensive analysis of the topology of bands and gaps beyond $E_1$ and $E_2$, aimed at identifying systematic trends as functions of $\tilde{d}$, $\tilde{\mathcal{A}}$, and $\tilde{B}$, is therefore demanding and lies beyond the scope of the present work.

Here, we instead present a representative example that illustrates the potential of honeycomb-based FM SkXs to enable frequency-multiplexed magnon transport through the coexistence of chiral TESs in distinct low-energy gaps. Specifically, we consider Model 2 ($\tilde{\mathcal{A}} = 0.103$) with $\tilde{d} = 0.93$, for which TESs occur simultaneously in the first and fourth magnon gaps. For these parameters, the SkX stabilizes at $\tilde{B}_{\min} \approx 4.8$ with a periodicity $w_0 = 6a$. The first six magnon bands are shown in Fig. 8(a), together with the corresponding Chern numbers for the lowest five bands. The Chern number $C_6$ is not relevant for the present discussion and is therefore omitted.

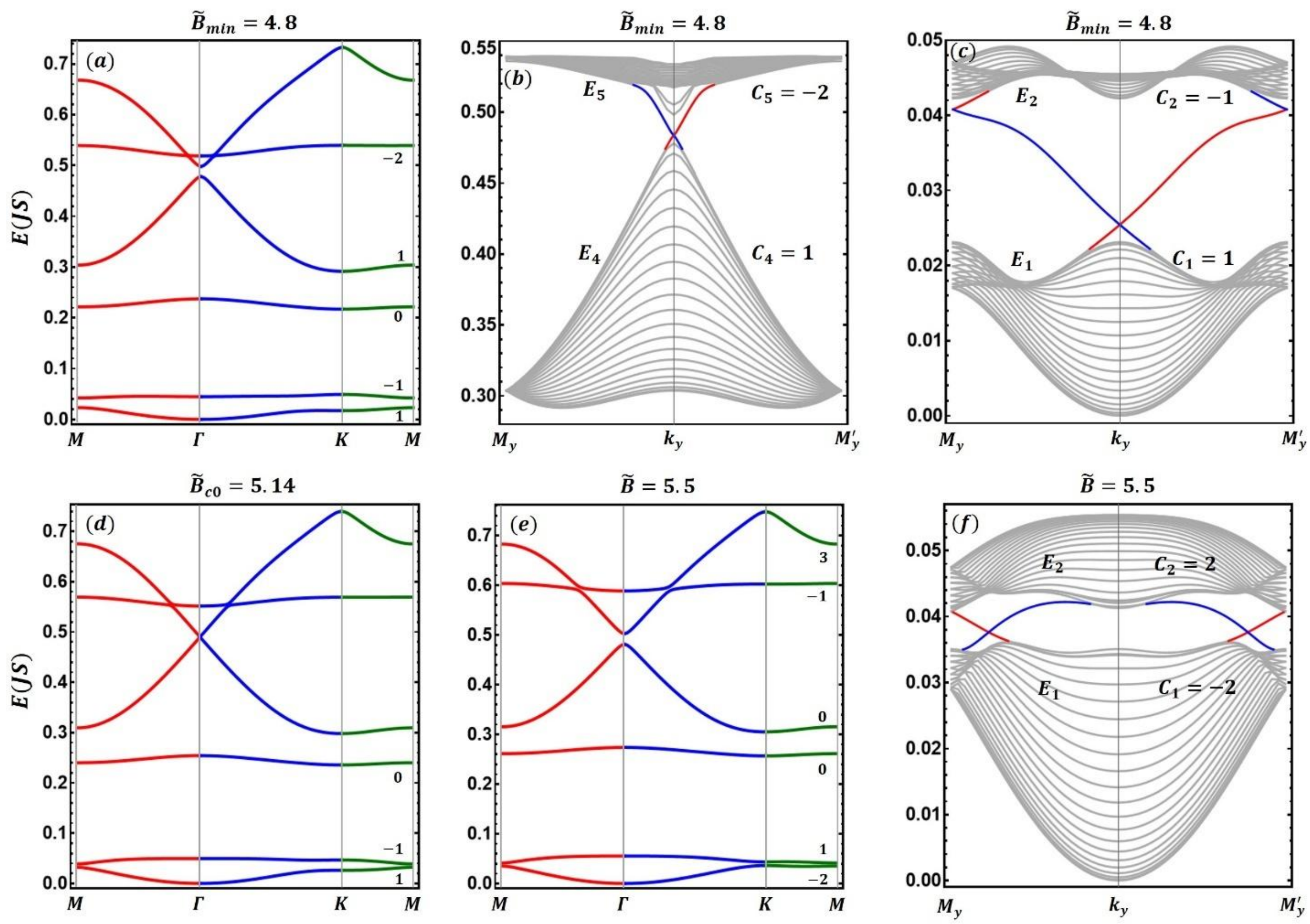

**Figure 8.** (a) Lowest six magnon bands for the densely packed SkX obtained at $\tilde{d} = 0.93$, $\tilde{\mathcal{A}} = 0.103$ (Model 2), $\tilde{B}_{\min} = 4.8$, and $w_0 = 6a$. The bands are plotted along the high-symmetry directions indicated in the inset of Fig. 2(a). The Chern numbers of the lowest five bands are listed. (b,c) Chiral TESs (red and blue curves) in the fourth (b) and first (c) magnon gaps at $\tilde{B}_{\min}$. The dispersions are plotted as functions of $k_y$ over the BZ segment illustrated in the inset of Fig. 3(a). (d) Lowest six magnon bands at the critical field $\tilde{B}_{c0} = 5.14$, where the fourth magnon gap closes. After the gap reopens for $\tilde{B} > \tilde{B}_{c0}$, the sum of the Chern numbers below this gap becomes zero (e), which eliminates the TESs. In contrast, TESs remain present in the first magnon gap (f) up to the critical field $\tilde{B}_{c2} \approx 5.7$ (see main text for details).

At $\tilde{B}_{\min}$, band $E_1$ corresponds to an ED mode, while $E_2$ is a CCW rotation mode, consistent with the results of Sec. 4.1. Band $E_3$ is topologically trivial and corresponds to a triangular distortion mode. In contrast, bands $E_4$ and $E_5$ are topological and correspond to a CCW rotation mode and a breathing mode, respectively. At this magnetic field, both the first and the fourth magnon gaps are topological, since $C_1 = 1$ and $\sum_{i=1}^{4} C_i = 1$. Consequently, each of these gaps hosts two chiral TESs, as illustrated in Fig. 8(c) for the first gap and Fig. 8(b) for the fourth gap.

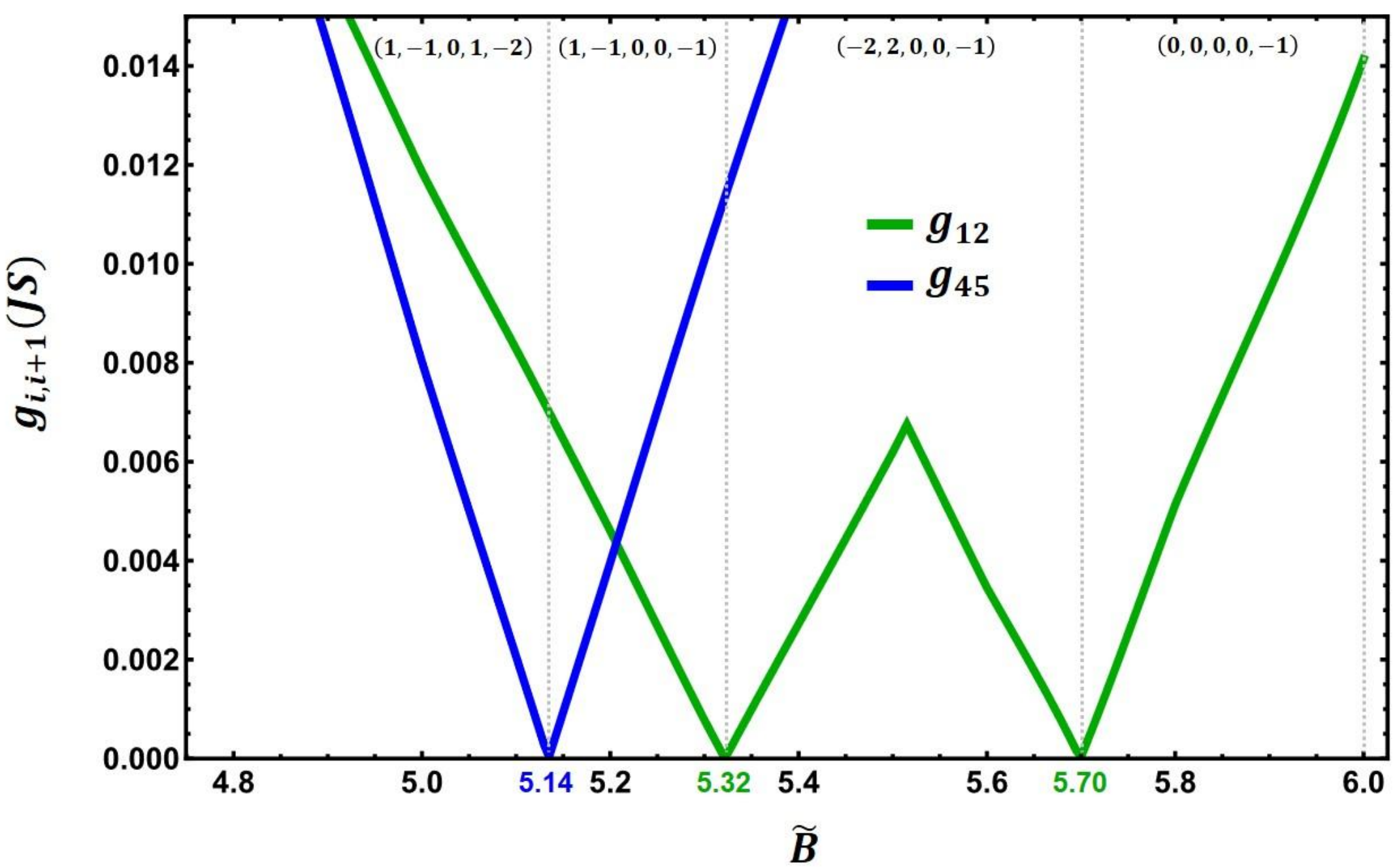

**Figure 9.** Minimal gaps $g_{12}$ and $g_{45}$, defined as the minimum values of $E_2 - E_1$ and $E_5 - E_4$, respectively, for a given magnetic field $\tilde{B}$ (see main text for details), plotted over the range $4.8 \leq \tilde{B} \leq 6$. The conditions $g_{12} = 0$ and $g_{45} = 0$ correspond to closures of the first and fourth magnon gaps, respectively. Within the displayed field range, the fourth gap closes once at $\tilde{B} \approx 5.14$, while the first gap closes twice, at $\tilde{B} \approx 5.32$ and $\tilde{B} \approx 5.7$. Each gap closure induces a TPT that modifies the Chern numbers of the bands bounding the gap. The corresponding Chern numbers $(C_1, \ldots, C_5)$ are indicated in the $\tilde{B}$ intervals delimited by the critical magnetic fields.

To track the evolution of these gaps with the magnetic field, we define $g_{i,i+1}(\tilde{B})$ as the minimum energy separation between two consecutive bands $E_i$ and $E_{i+1}$. This quantity is given by the minimum of $E_{i+1}(\boldsymbol{k}, \tilde{B}) - E_i(\boldsymbol{k}, \tilde{B})$ over the BZ at fixed $\tilde{B}$. A vanishing $g_{i,i+1}$ signals a gap closure and hence a TPT. In Fig. 9, we plot $g_{1,2}(\tilde{B})$ and $g_{4,5}(\tilde{B})$ as functions of the magnetic field in the range $4.8 \leq \tilde{B} \leq 6$. The results show that the fourth gap closes at $\tilde{B}_{c0} \approx 5.14$ (see also Fig. 8(d)), while the first gap closes twice, at $\tilde{B}_{c1} \approx 5.32$ and $\tilde{B}_{c2} \approx 5.70$, in agreement with the analysis of Sec. 4.1. Each gap closure induces a TPT, and the corresponding Chern numbers $(C_1, \dots, C_5)$ for the different phases are indicated in Fig. 9.

An analysis of these Chern numbers shows that chiral TESs coexist in the first and fourth magnon gaps over the magnetic field range $\tilde{B}_{\min} \leq \tilde{B} < \tilde{B}_{c0}$. This regime therefore offers two spectrally separated, low-energy edge-state channels that could enable frequency-multiplexed chiral magnon transport. At $\tilde{B}_{c0}$, the TPT in the fourth gap inverts the CCW and breathing modes, changing $(C_4, C_5)$ from $(1, -2)$ to $(0, -1)$ after the gap reopens (Figs. 9 and 8(e)). As a result, the fourth gap becomes topologically trivial, since $\sum_{i=1}^{4} C_i = 0$ for $\tilde{B}_{c0} < \tilde{B} \leq 6$, and its TESs disappear. In contrast, the TESs in the first gap persist up to $\tilde{B}_{c2}$, and their number is doubled by the transition at $\tilde{B}_{c1}$ (Fig. 8(f)), consistent with the discussion in Sec. 4.1. The corresponding evolution of the modal characters of bands $E_1$ and $E_2$ has also been discussed in Sec. 4.1.

Before closing this section, we note that topological edge states have previously been reported in the CCW–breathing gap of triangular-lattice SkXs [4], which bears some resemblance to the results presented here. However, several significant differences exist. First, the CCW–breathing gap corresponds to the third magnon gap in the triangular case, rather than the fourth gap in the honeycomb case. Second, the associated TPT in the triangular lattice changes the Chern numbers $(C_3, C_4)$ from $(1,1)$ to $(0,1)$, which differs from the redistribution of topological charge observed here. Most importantly, in the triangular case the lowest two bands $E_1$ and $E_2$ are topologically trivial, precluding the possibility of frequency-multiplexed chiral magnon transport.

## 5. Discussion

### 5.1 Remarks on commensurability and parameter limits.

Periodic SkXs require the lattice periodicity $w_0$ to be an integer multiple of the spin lattice constant $a$. However, due to the random nucleation of $\tilde{d}$-induced skyrmions, the normalized periodicity $\tilde{w}_0 = w_0/a$ obtained from the Vampire simulations is generally noninteger and must be rounded to construct a periodic SkX. In rare cases, $\tilde{w}_0$ approaches a half-integer value, which renders this rounding procedure ambiguous.

Two such cases were identified, namely $(\tilde{\mathcal{A}} = 0.15, \tilde{d} = 1)$ and $(\tilde{\mathcal{A}} = 0.08, \tilde{d} = 0.55)$, for which $\tilde{w}_0$ lies slightly below 5.5 and 8.5, respectively. Adopting $w_0 = 5a$ and $w_0 = 8a$ in these two cases does not yield TESs in the first magnon gap, whereas TESs do appear in the vicinity of these special $\tilde{d}$ values. Although no additional cases were detected in our simulations, the existence of a few other exceptional $\tilde{d}$ values cannot be excluded within the $\tilde{d}$ ranges shown in Fig. 5.

We now comment on the implications of our results for larger values of the SIMA $\tilde{\mathcal{A}}$. In the present work, since $\tilde{d}_{\mathrm{th}} = 0.55$ and $\tilde{d}_{\max} = 1.2$ are already relatively large at $\tilde{\mathcal{A}} = 0.15$ compared to typical experimentally accessible values, we adopt $\tilde{\mathcal{A}} = 0.15$ as a practical upper bound. We note, however, that substantially larger NN DMI values, such as $\tilde{d} = 2.16$, have been considered in recent theoretical studies [7,9] and physically justified through mechanisms that effectively tune the exchange interaction $J$ [54–56]. Exploring the extension of our analysis to higher values of $\tilde{\mathcal{A}}$ therefore remains an interesting direction for future work.

While our study is restricted to $\tilde{\mathcal{A}} \leq 0.15$, the trends identified here allow for qualitative predictions at larger anisotropy. In particular, we expect the NN DMI window supporting first-gap TESs, $R_{\mathrm{DMI}}$, to widen with increasing $\tilde{\mathcal{A}}$. At the same time, we emphasize that, at sufficiently large $\tilde{\mathcal{A}}$, one cannot exclude the possibility that triangular-based SkXs may also host TESs in the lowest magnon gap, although this would likely occur over a significantly narrower $\tilde{d}$ range than in the honeycomb case.

### 5.2 Symmetry-based interpretation for the emergence of TESs in the first gap.

We have shown that the lowest magnon gap in honeycomb-based SkXs is topological over a region of the $(\tilde{\mathcal{A}}, \tilde{d})$ parameter plane. This behavior is absent in triangular-based SkXs, at least within the range of $\tilde{\mathcal{A}}$ considered in this work. As discussed in Sec. 4, this qualitative difference does not originate from the NNN DMI $\widetilde{D}$ within the experimentally relevant values explored here, for which its effect on the magnon topology in the SkX phase is found to be negligible. Moreover, it cannot be attributed solely to differences in the modal characters of bands $E_1$ and $E_2$, since a given mode may be either topological or trivial depending on the magnetic field. Instead, we argue that the emergence of TESs in the first magnon gap is rooted in a local symmetry of the honeycomb-based magnon Hamiltonian in momentum space. This symmetry is absent in the triangular case, at least for the parameter ranges considered here.

To set the stage, it is instructive to examine the structural differences between triangular-based and honeycomb-based skyrmions, as illustrated in Fig. 10. The honeycomb lattice has a two-site basis and therefore contains twice as many spins per magnetic unit cell as the triangular lattice. In addition, each site in the honeycomb lattice has three NN, compared to six in the triangular lattice.

These differences lead to distinct skyrmion geometries and to different NN exchange and DMI interaction pathways, which in turn affect the structure of the resulting magnon Hamiltonian. Indeed, the magnon Hamiltonian for the honeycomb-based SkX, derived in the Supplementary Material, has a different structure from that obtained previously for triangular lattices [4,5]. Moreover, these distinct interaction pathways result in different values of $\tilde{B}_{\min}$ and $w_0$ in the sLLG simulations for the triangular and honeycomb cases, even for identical values of $(\tilde{\mathcal{A}}, \tilde{d})$. For example, as shown in Sec. 4.1, for $\tilde{\mathcal{A}} = 0.15$ and $\tilde{d} = 1.2$ the triangular-based SkX stabilizes at higher $\tilde{B}_{\min}$ and exhibits a larger skyrmion periodicity $w_0$ than its honeycomb counterpart. Both $\tilde{B}_{\min}$ and $w_0$ enter directly into the magnon Hamiltonian and affect the results.

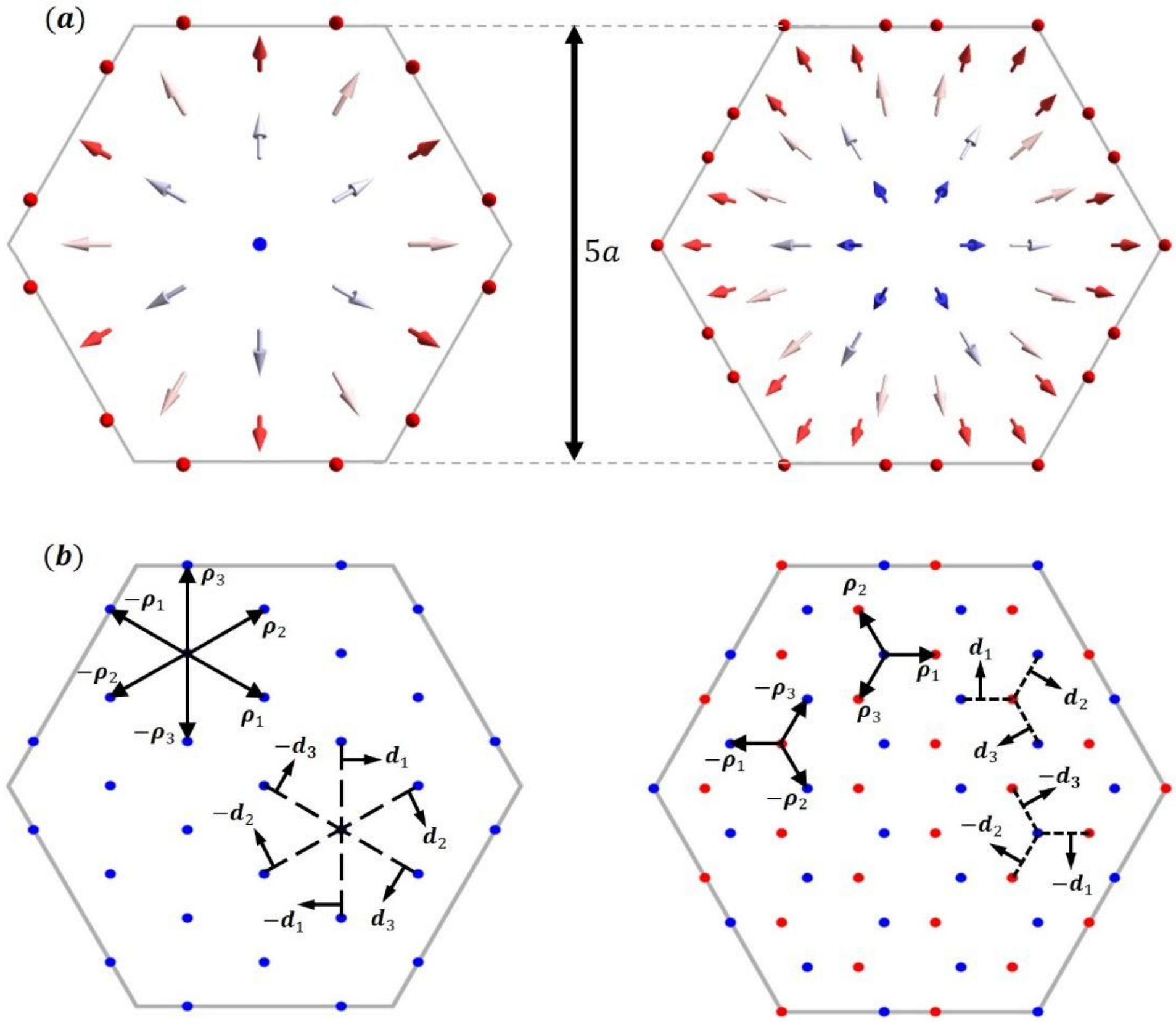

**Figure 10.** (a) Schematic representations of skyrmions with periodicity $w_0 = 5a$ formed on triangular (left) and honeycomb (right) spin lattices. The skyrmions encompass 25 and 50 inequivalent spins in the triangular and honeycomb cases, respectively. Here, inequivalent spins are defined as those that cannot be mapped onto each other by translations of the SkX lattice vector. (b) NN displacement vectors $\boldsymbol{\rho}_j$ and NN DMI vectors $\boldsymbol{d}_j$ for the triangular (left) and honeycomb (right) lattices. In the honeycomb case, the red and blue sites denote the conventional $A$ and $B$ sublattices. Each site in the triangular lattice interacts via exchange and DMI with six NNs, whereas in the honeycomb

lattice each site has only three NNs. Although the vectors $\boldsymbol{\rho}_j$ differ in the two lattices, the same notation is used for simplicity.

We now consider a SkX corresponding to one of the $(\tilde{\mathcal{A}}, \tilde{d})$ parameter sets shown in Fig. 5, initially subjected to a magnetic field $\tilde{B} > \tilde{B}_{c2}$. In this regime, the two lowest magnon bands $E_1$ and $E_2$ are topologically trivial. Upon decreasing the magnetic field toward $\tilde{B}_{\min}$, at least one topological phase transition occurs at $\tilde{B}_{c2}$. This transition is associated with a closure of the first magnon gap at the $K$ point of the BZ and leads to the emergence of TESs in the first gap after it reopens. At the gap-closing point, bands $E_1$ and $E_2$ become locally degenerate in momentum space, indicating the presence of a symmetry that enforces this degeneracy. The absence of analogous TPTs in the triangular case suggests that this symmetry is not realized there.

To identify the symmetry underlying the $E_1$–$E_2$ gap closures in the honeycomb case, we recall the momentum space magnon Hamiltonian introduced in Eq. (3). Upon diagonalization via a Bogoliubov transformation, the Hamiltonian takes the form,

$$\mathcal{H} = \sum_{\boldsymbol{k},\mu} E_\mu(\boldsymbol{k}) \left(\alpha^+_{\boldsymbol{k},\mu}\alpha_{\boldsymbol{k},\mu} + \tfrac{1}{2}\right) \tag{4}$$

where $E_\mu(\boldsymbol{k})$ denotes the energy of the $\mu$-th magnon band at momentum $\boldsymbol{k}$. Focusing on two consecutive bands $\nu$ and $\nu + 1$, one may define an effective $2 \times 2$ Hamiltonian,

$$\mathcal{H}_{\nu,\nu+1} = E_\nu(\boldsymbol{k})\, \alpha^+_{\boldsymbol{k},\nu}\alpha_{\boldsymbol{k},\nu} + E_{\nu+1}(\boldsymbol{k})\, \alpha^+_{\boldsymbol{k},\nu+1}\alpha_{\boldsymbol{k},\nu+1} \tag{5}$$

which acts on the vector space $\mathcal{V}_{\nu,\nu+1}$ spanned by the corresponding eigenstates. Since we are interested in the first magnon gap, we set $\nu = 1$ in the following.

For $\nu = 1$, Eq. (5) can be rewritten as

$$\mathcal{H}_{12} = f_+(\boldsymbol{k}) \left(\alpha^+_{\boldsymbol{k},1}\alpha_{\boldsymbol{k},1} + \alpha^+_{\boldsymbol{k},2}\alpha_{\boldsymbol{k},2}\right) + f_-(\boldsymbol{k}) \left(\alpha^+_{\boldsymbol{k},1}\alpha_{\boldsymbol{k},1} - \alpha^+_{\boldsymbol{k},2}\alpha_{\boldsymbol{k},2}\right) \tag{6}$$

with $f_{\pm}(\boldsymbol{k}) = \frac{1}{2}[E_1(\boldsymbol{k}) \pm E_2(\boldsymbol{k})]$.

We now introduce the operators $J_z = \frac{1}{2}\hbar\left(\alpha^{+}_{\boldsymbol{k},1}\alpha_{\boldsymbol{k},1} - \alpha^{+}_{\boldsymbol{k},2}\alpha_{\boldsymbol{k},2}\right)$, $J_+ = \hbar\alpha^{+}_{\boldsymbol{k},1}\alpha_{\boldsymbol{k},2}$, and $J_- = \hbar\alpha^{+}_{\boldsymbol{k},2}\alpha_{\boldsymbol{k},1}$. Using the bosonic commutation relations $\left[\alpha_{\boldsymbol{k},\nu}, \alpha^{+}_{\boldsymbol{k}',\nu'}\right] = \delta_{\boldsymbol{k},\boldsymbol{k}'}\delta_{\nu,\nu'}$ for the Bogoliubov operators, one readily verifies that $\{J_z,\ J_+,\ J_-\}$ satisfy the standard angular momentum algebra, specifically $[J_z, J_\pm] = \pm\hbar\, J_\pm$ and $[J_+, J_-] = 2\hbar\, J_z$. This hidden angular momentum operator, denoted $\boldsymbol{J}$, generates a rotation operator $\mathcal{D}_{\boldsymbol{n}}(\theta) = \exp\left(\frac{i}{\hbar}\ \theta\boldsymbol{J}.\boldsymbol{n}\right)$. $\mathcal{D}_{\boldsymbol{n}}(\theta)$ acts within the 2D space $\mathcal{V}_{1,2}$, where $\boldsymbol{n}$ is a unit vector and $\theta$ is a continuous parameter.

The Hamiltonian $\mathcal{H}_{12}$ can be expressed in terms of $J^2$ and $J_z$. From Eq. 6 and the definition of $J_z$, we can write

$$\mathcal{H}_{12} = f_+(\boldsymbol{k})\, \mathcal{N} + \frac{2}{\hbar} f_-(\boldsymbol{k})\, J_z \tag{7}$$

where $\mathcal{N} = \alpha^{+}_{\boldsymbol{k},1}\alpha_{\boldsymbol{k},1} + \alpha^{+}_{\boldsymbol{k},2}\alpha_{\boldsymbol{k},2}$. Now substituting the Bogoliubov operators in the identity $J^2 = J_-J_+ + J_z^2 + \hbar J_z$, we arrive at the identity

$$\mathcal{N}^2 + 2\mathcal{N} - \left(\frac{2}{\hbar}\, J\right)^2 = 0 \tag{8}$$

Eq. 8 states that the operators $J^2$ and $\mathcal{N}$ commute, and the equation can be easily solved to get $\mathcal{N} = \sqrt{1 + \left(\frac{2}{\hbar}\, J\right)^2} - 1$. Substituting in Eq. 7 gives,

$$\mathcal{H}_{1,2} = f_+(\boldsymbol{k}) \left[\sqrt{1 + \left(\frac{2}{\hbar}\, J\right)^2} - 1\right] + \frac{2}{\hbar} f_-(\boldsymbol{k})\, J_z \tag{9}$$

This representation makes the symmetry properties of $\mathcal{H}_{12}$ explicit. When the gap between bands $E_1$ and $E_2$ closes for a specific combination of the parameters ($w_0$, $\tilde{B}$, $\tilde{d}$, $\tilde{D}$, $\boldsymbol{k}$ and $\tilde{\mathcal{A}}$), one has $E_1 = E_2$ and therefore $f_-(\boldsymbol{k}) = 0$. In this case, the Hamiltonian reduces to

$$\mathcal{H}_{1,2} = f_+(\boldsymbol{k}) \left[ \sqrt{1 + \left(\frac{2}{\hbar} J\right)^2} - 1 \right] \tag{10}$$

which is invariant under arbitrary rotations $\mathcal{D}_{\boldsymbol{n}}(\theta)$, since $J^2$ commutes with $\boldsymbol{J}$. The resulting degeneracy at the gap-closing point is therefore protected by this rotational symmetry. We find that the honeycomb-based SkX Hamiltonian supports this symmetry at specific parameter values, whereas the corresponding symmetry is absent in triangular-based SkXs within the parameter ranges considered in this work.

## 6. Conclusion

We have presented a theoretical study of magnon band topology and TESs in Néel-type FM SkXs stabilized on a 2D honeycomb spin lattice. We demonstrated that the honeycomb lattice geometry enables the emergence of chiral TESs in the first bulk magnon gap over significant ranges of SIMA and NN DMI. This behavior is absent in FM SkXs on triangular lattices within the same parameter regime.

We showed that these first-gap TESs undergo magnetic-field-driven topological phase transitions that modify their number and ultimately trivialize them at higher fields. By systematically varying the magnetic anisotropy, we identified a lower bound below which first-gap TESs no longer emerge. Within this framework, honeycomb-based van der Waals magnets with sizable anisotropy, such as monolayer $CrI_3$, can support low-energy chiral edge modes, whereas systems with much smaller anisotropy, such as $CrBr_3$, do not.

Beyond the lowest-energy sector, we demonstrated that honeycomb-based SkXs can simultaneously host chiral edge states in the first and higher gaps, over finite magnetic-field ranges. This coexistence illustrates the potential of these systems for frequency-multiplexed chiral magnon transport. We further established that moderate values of NNN DMI, while crucial in collinear honeycomb ferromagnets, do not significantly affect the magnon topology in the SkX phase.

Finally, we provided a symmetry-based interpretation for the emergence of first-gap TESs in the honeycomb case. We showed that gap closures between the lowest magnon bands are associated with an emergent local symmetry of an effective two-band magnon Hamiltonian at specific momenta and parameter values. This symmetry is supported by the honeycomb lattice structure and the noncollinear skyrmion background, but is absent in triangular-based SkXs within the parameter ranges of our study.

## Acknowledgments

Part of the numerical calculations was performed using the Phoenix High Performance Computing facility at the American University of the Middle East (AUM), Kuwait.